\documentclass[british,english,aps,prb,reprint]{revtex4-1}
\usepackage[T1]{fontenc}
\usepackage[latin9]{inputenc}
\usepackage{color}
\usepackage{babel}
\usepackage{array}
\usepackage{booktabs}
\usepackage{amsmath}
\usepackage{graphicx}
\usepackage[unicode=true,
 bookmarks=true,bookmarksnumbered=false,bookmarksopen=false,
 breaklinks=false,pdfborder={0 0 1},backref=false,colorlinks=true]
 {hyperref}
\hypersetup{
 linkcolor=red, citecolor=blue, urlcolor=green, pdfstartview=XYZ, plainpages=false, pdfpagelabels}

\makeatletter

\providecommand{\tabularnewline}{\\}

\@ifundefined{date}{}{\date{}}
\PassOptionsToPackage{square,comma}{natbib}
\RequirePackage{natbib}

\usepackage{etoolbox}

\makeatother

\begin{document}
\title{Thermodynamic stability and vibrational anharmonicity of black phosphorene
- beyond quasi-harmonic analysis}
\author{\selectlanguage{british}%
P. Anees}
\email{anees@igcar.gov.in}

\affiliation{Materials Physics Division, Indira Gandhi Centre for Atomic Research,
Kalpakkam 603102, Tamil Nadu, India}
\begin{abstract}
Thermodynamic stability and vibrational anharmonicity of single layer
black phosphorene (SLBP) are studied using a spectral energy density
(SED) method. Thermal stability of SLBP sheet is analyzed by computing
phonon dispersion at 300 K, which shows that SLBP sheet is dynamically
stable at finite temperature and survives the crumpling transition.
Temperature evolution of all zone center optic phonon modes are extracted,
including experimentally forbidden IR and Raman active modes. Mode
resolved phonon frequencies of optic modes shows significant deviation
from quasi-harmonic prediction, which is ascribed to the effects of
inclusion of higher order phonon-phonon scattering processes. Further,
temperature sensitivity of each mode is analyzed by computing their
first order temperature co-efficient ($\chi$). The quasi-harmonic
{\normalsize{}$\chi$} values are one order magnitude smaller than
the SED and experimental values; which again substantiate that quasi-harmonic
methods are inadequate, and a full anaharmonic analysis is essential
to explain structure and dynamics of SLBP at finite temperatures.
\end{abstract}
\maketitle
The search for 2D materials with novel, structural, electronic and
thermal properties has geared up in past few years \citep{Zhou2019}.
Single layer of black phosphorous, known as black phosphorene (SLBP)
is a special member in 2D family due to its intriguing structural
and electronic properties \citep{Li2014,Churchill2014,Xia2014}. The
band gap ($\approx$ 1.5 eV) \citep{doi:10.1021/nn501226z} and carrier
mobility ($\approx$ 1000 cm\textsuperscript{2} V\textsuperscript{-1}
s\textsuperscript{-1})\citep{Li2014} of SLBP make it a promising
candidate for nano \citep{doi:10.1021/nn501226z,Li2014,Xia2014} and
opto-electronic device fabrications \citep{Churchill2014,Xia2014}.
SLBP also found its application in photo-transistors \citep{doi:10.1021/nn501226z,Li2014,Xia2014}\textsuperscript{,}\citep{Qin2014,PhysRevB.90.081408,PhysRevB.90.085433},
thermo-electric devices \citep{Qin2014,PhysRevB.90.085433,doi:10.1021/nl502865s}
and gas sensors \citep{Suvansinpan_2016}. 

Being a technologically promising material in many industries, it
is essential to have knowledge on its vibrational and thermal properties.
The vibrational dynamics of mono and few-layer phosphorene have been
extensively studied using Raman scattering techniques\citep{doi:10.1063/1.4894273,doi:10.1063/1.4928931,doi:10.1021/acs.jpcc.6b01468}
As per group theoretical analysis \citep{doi:10.1002/adfm.201404294},
there are six Raman active modes but only three of them have been
detected due to the constraints imposed by Raman scattering geometry.
Similarly the temperature evolution of IR active modes are also not
reported in literature. In Raman spectroscopy, factors like substrate
effect, thickness, laser power and morphology etc., will alter the
intrinsic phonon transport in SLBP \citep{doi:10.1021/nn503893j,doi:10.1021/acs.jpclett.5b00043,doi:10.1063/1.4928931}. 

Under such circumstances atomistic simulations will help to predict
the intrinsic material properties.\textit{ Ab initio} simulation have
been employed to study the lattice dynamics properties of SLBP \citep{PhysRevB.92.081408,doi:10.1002/adfm.201404294,doi:10.1021/acs.jpcc.6b01468}.
Also, strain evolution of phonon modes are studied in detail within
the \textit{ab inito} frame work \citep{Wang2015,PhysRevB.92.081408,doi:10.1063/1.4894273}.
Aierken \textit{et al }\citep{PhysRevB.92.081408}\textcolor{green}{{}
}\textcolor{black}{used} the quasi-harmonic approximation (QHA) to
understand the thermal and vibrational properties. The authors warned
the usage of quasi-harmonic approach at high temperature due to the
emergence of strong phonon-phonon interaction, which is not incorporated
in QHA. Motivated from the above facts, the objective of the present
study is set as, analyze the thermodynamic stability and intrinsic
vibrational anharmonicity of all optic phonon modes in SLBP with full
anharmonicity of effective interaction, irrespective of present experimental
and theoretical constraints.

To compute the finite temperature structural and vibrational properties
a spectral energy density (SED) method is used in conjunction with
MD simulation \citep{P.Anees2015}. In SED method, anharmonicities
of all orders are naturally incorporated, hence it would bring out
the intrinsic phonon transport properties. A detailed theoretical
derivations of SED can be found in our previous paper \citep{P.Anees2015}.
For the sake of readability, a brief description of SED method is
given below. The SED method works in two sequences, firstly a lattice
dynamics (LD) calculation is done to obtain the polarization vector
$\{\arraycolsep=1.4pt
\global\long\def\arraystretch{0.5}%
e_{\alpha}\bigg(\kappa\,\bigg\vert\begin{array}{c}
\overline{q}\\
j
\end{array}\bigg),\,\alpha=x,y,z;\,\kappa=1.....N_{b}\}$ of $\binom{\bar{q}}{j}^{th}$normal mode of vibration (where, $\bar{q}$
= wave vector, $j$ = mode index \& $N_{b}$= number of basis atoms).
Secondly, MD simulations are performed at desired temperature to obtain
the velocities, $\arraycolsep=1.4pt
\global\long\def\arraystretch{0.5}%
v_{\alpha}\bigg(\begin{array}{c}
l\\
\kappa
\end{array}\bigg\vert\,t\bigg)$ of $\binom{l}{\kappa}^{th}$ atom (where $l$ is the cell-index)
at time t. Further, project these velocities onto $\binom{\bar{q}}{j}^{th}$mode
of vibration, and thus define a quantity $\arraycolsep=1.4pt
\global\long\def\arraystretch{0.5}%
\psi\bigg(\begin{array}{c}
\bar{q}\\
j
\end{array}\bigg\vert\,t\bigg)$ as, 

\[
\arraycolsep=1.4pt
\global\long\def\arraystretch{0.5}%
\psi\bigg(\begin{array}{c}
\bar{q}\\
j
\end{array}\bigg\vert\,t\bigg)=\sum_{\alpha,\kappa}\,\Biggl(\begin{alignedat}[b]{1}\sum_{l} & \,\end{alignedat}
v_{\alpha}\bigg(\begin{array}{c}
l\\
\kappa
\end{array}\bigg\vert\,t\bigg)\exp\Bigl(-i\bar{q}.\bar{r}(l)\Bigr)\Biggr)\,e_{\alpha}^{*}\bigg(\kappa\,\bigg\vert\begin{array}{c}
\overline{q}\\
j
\end{array}\bigg)
\]

The Fourier transform and then power spectrum $\Biggl|\arraycolsep=1.4pt
\global\long\def\arraystretch{0.5}%
\tilde{\psi}\bigg(\begin{array}{c}
\bar{q}\\
j
\end{array}\bigg\vert\,\omega\bigg)\Biggr|^{2}$will yield the mode resolved phonon frequencies and linewidths of
all phonon modes at any $\overline{q}$ point in the Brillouine zone.

\begin{figure*}
\centering{}\includegraphics[scale=0.31]{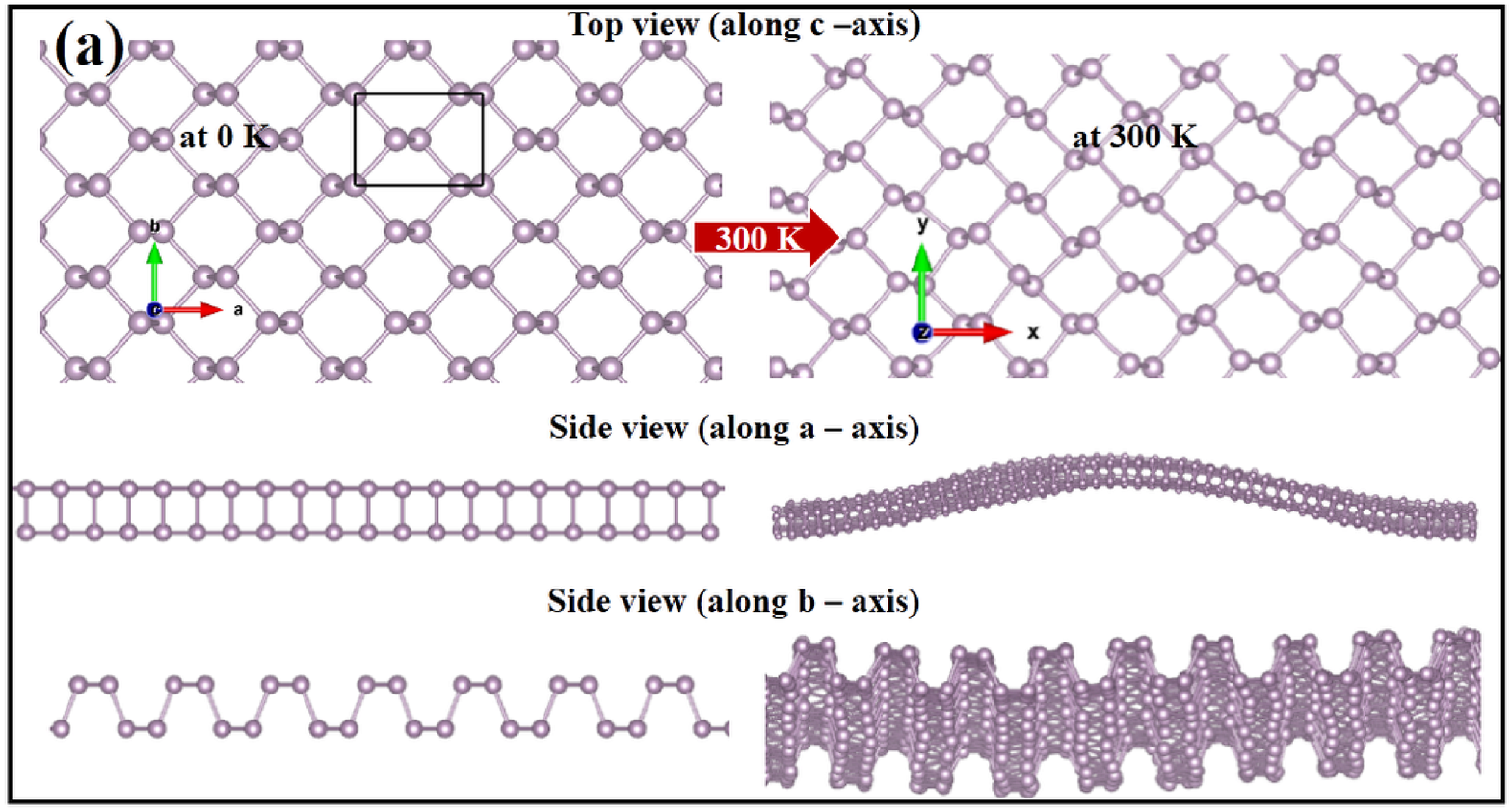}\includegraphics[scale=0.26]{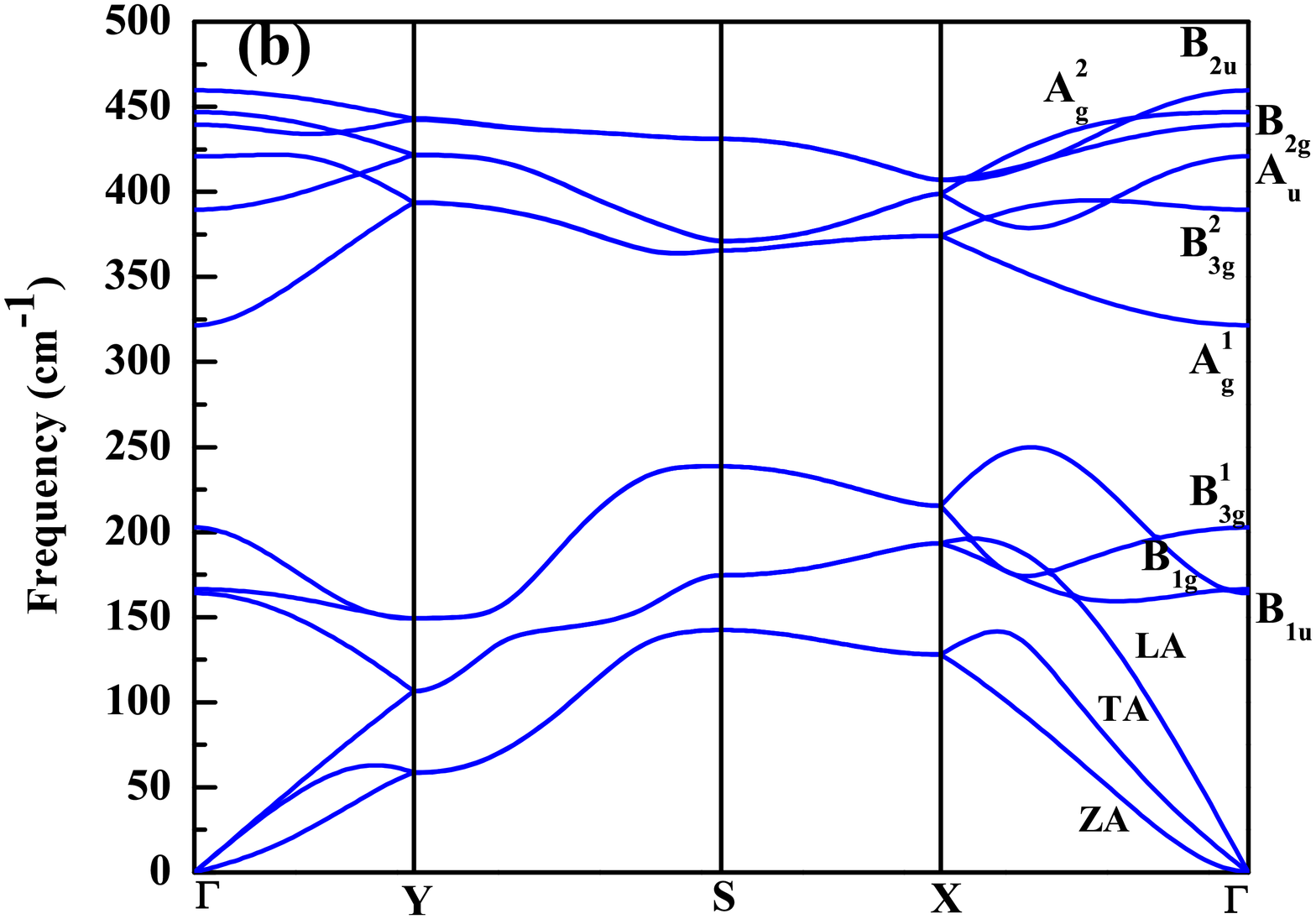}\caption{(a) The atomistic structure of SLBP sheet at 0K and 300 K. The sheet
is corrugated at 300 K due to the formation of ripples at finite temperatures;
(b) phonon dispersion curve at T = 0 K\label{fig:structure-at-0K}.}
\end{figure*}
In the present study, MD simulation are performed using LAMMPS package
\citep{Plimpton19951}. The phosphorous-phosphorous interaction in
SLBP is defined using Stillinger-Weber (SW) potential parametrized
by Jiang \textit{et al }\citep{C4NR07341J}. The SW potential describes
the non-linear properties of covalent crystals, reasonably, accurately
\citep{ABRAHAM1989L125}, which makes the prediction of higher order
phonon transport more accurate in SLBP \citep{C5NR03577E}. Periodic
boundary conditions are employed in all three direction, and a vacuum
separation of 20 \AA \, is provided along \textbf{c}-axis to eliminate
the un-physical interactions between the periodic images. Further,
to \textcolor{black}{get rid}\textbf{\textcolor{black}{{} }}\textcolor{black}{of}
the residual stresses, the system is relaxed in conjugate gradient
algorithm. The lattice parameters of relaxed structures are \textbf{a}
= 4.596 \AA \, \&\,\textbf{ b} = 3.278 \AA, which shows an excellent
agreement with earlier studies\citep{doi:10.1021/acs.jpclett.5b00522,doi:10.1063/1.4928931}.
Phonon frequencies and polarizations are computed under harmonic approximation
(T = 0 K) using a combination \citep{P.Anees2015} of LAMMPS and PHONOPY
\citep{phonopy}. To compute the finite temperature properties a simulation
cell of size 50x50x1 (10000 atoms) is adapted, and the system is equilbrated
in isobaric-isothermal (NPT) ensemble for 0.5 ns. Once ensuring the
equilbration \& thermalization, the NPT ensemble is unfixed and coupled
to NVE ensemble; further the velocities of each atoms are collected
in a predefined interval of 5 fs, and the total simulations time is
3.2 ns.

The unit-cell of SLBP is a rectangle containing four basis atoms and
belongs to $pnma\,(D_{2h}^{7})$ space group (Figure \ref{fig:structure-at-0K}).
Similar to graphene and 2D h-BN, SLBP also posses a honeycomb lattice
structure, covalentely bonded P atoms forms a puckered like structure.
SLBP unitcell contains 4 basis atom, hence there are 12 modes of vibration
(3 acoustic (A) + 9 optic (O)). Figure \ref{fig:structure-at-0K}
shows the phonon dispersion (at T =0 K) of SLBP. The acoustic modes
are labeled based on their polarization (L-longitudinal, T-transverse
and Z-out of plane) and the optic modes are designated as per group
theoretical notation at $\Gamma$ point. The absence of imaginary
modes in phonon dispersion ensures the structural stability of SLBP.
The out of plane acoustic (ZA) mode shows a $q^{2}$ behaviour near
the long wavelength limit, which is a fingerprint of typical 2D system
\citep{Lifshitz1952}. The slope and thus group velocities of LA and
TA branch is relatively small along armchair ($\Gamma-Y$) with respect
to zig-zag ($\Gamma-X$) direction, also the high frequency optic
mode frequencies are more flat along the armchair direction in comparison
with zig-zag and it is in accordance with previous observation \citep{doi:10.1002/adfm.201404294}. 

According to group theoretical analysis \citep{doi:10.1002/adfm.201404294},
among nine optic modes, two are IR active ($B_{1u}$ and $B_{2u}$),
six are Raman active ($A_{g}^{1}$, $A_{g}^{2}$, $B_{1g}$, $B_{2g}$,
$B_{3g}^{1}$, \textbf{$B_{3g}^{2}$}) and one is inactive ($A_{u}$).
Though there are six Raman active modes, only three ($A_{g}^{1}$,
$B_{2g}$, $A_{g}^{2}$) of them have been reported experimentally\citep{doi:10.1002/adfm.201404294}.
The LD $A_{g}^{1}$ mode frequency is 321.59 cm\textsuperscript{-1},
which is underestimated by 10.8 \% in comparison with Raman data (360.61
cm\textsuperscript{-1})\citep{doi:10.1021/nn501226z}. The $B_{2g}$
mode predicted by LD (439.55 cm\textsuperscript{-1}) shows an excellent
agreement with Raman measurement (438.90 cm\textsuperscript{-1})\citep{doi:10.1021/nn501226z}.
The difference is only 0.1 \% which is negligibly small. Similarly,
the LD $A_{g}^{2}$ mode frequency (446.90 cm\textsuperscript{-1})
is overestimated by 4.7 \% in comparison with Raman values (469.00
cm\textsuperscript{-1}) \citep{doi:10.1021/nn501226z}. The agreement
between the LD and experiment is satisfactory; although the present
SW potential is the best among available in literature for predicting
phonon spectra and their non-linear interaction, it needs further
refinement to make the prediction more accurate. 

\begin{figure*}
\begin{centering}
\includegraphics[width=7.5cm,height=7cm]{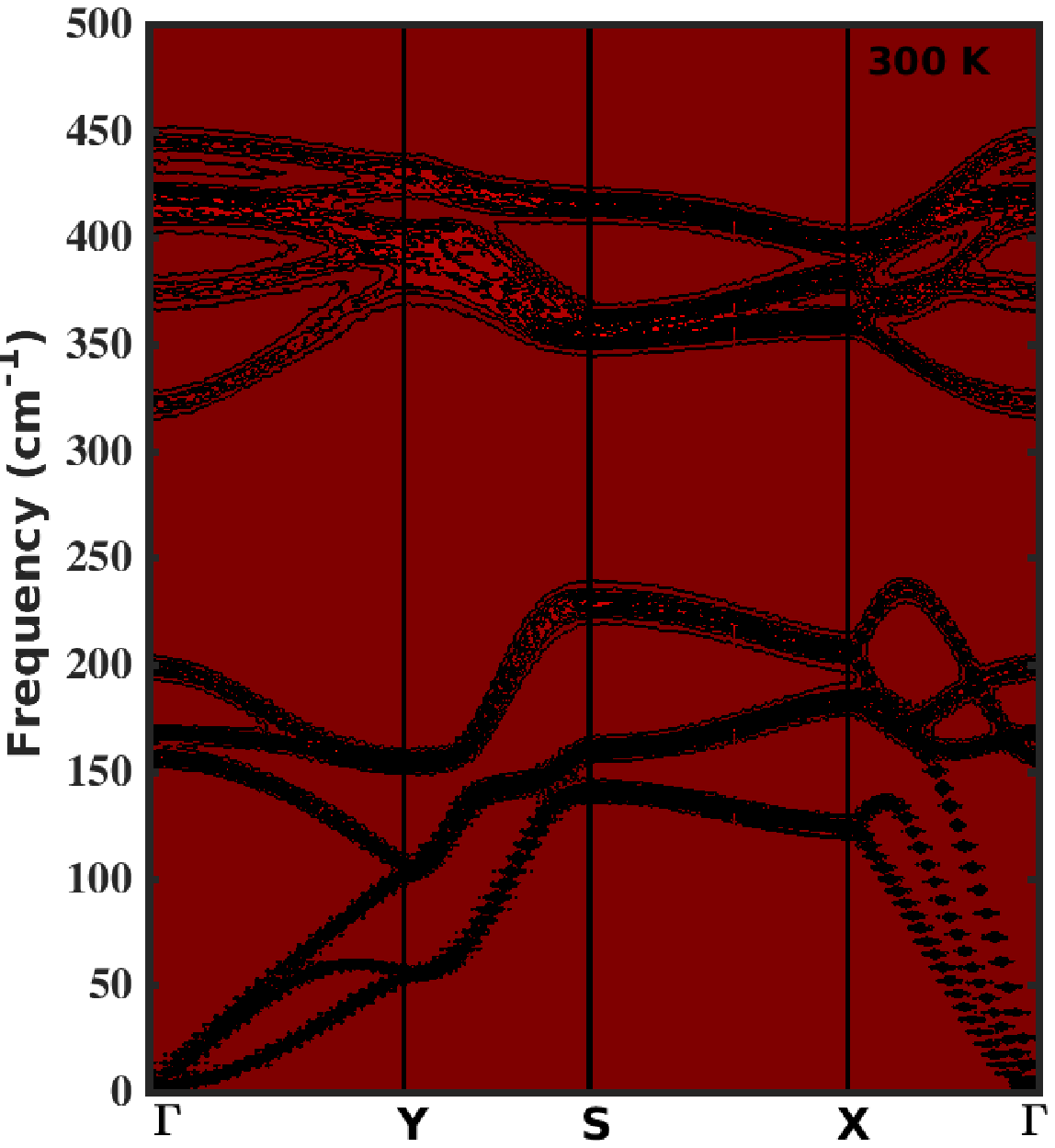}\includegraphics[width=7cm,height=6.5cm]{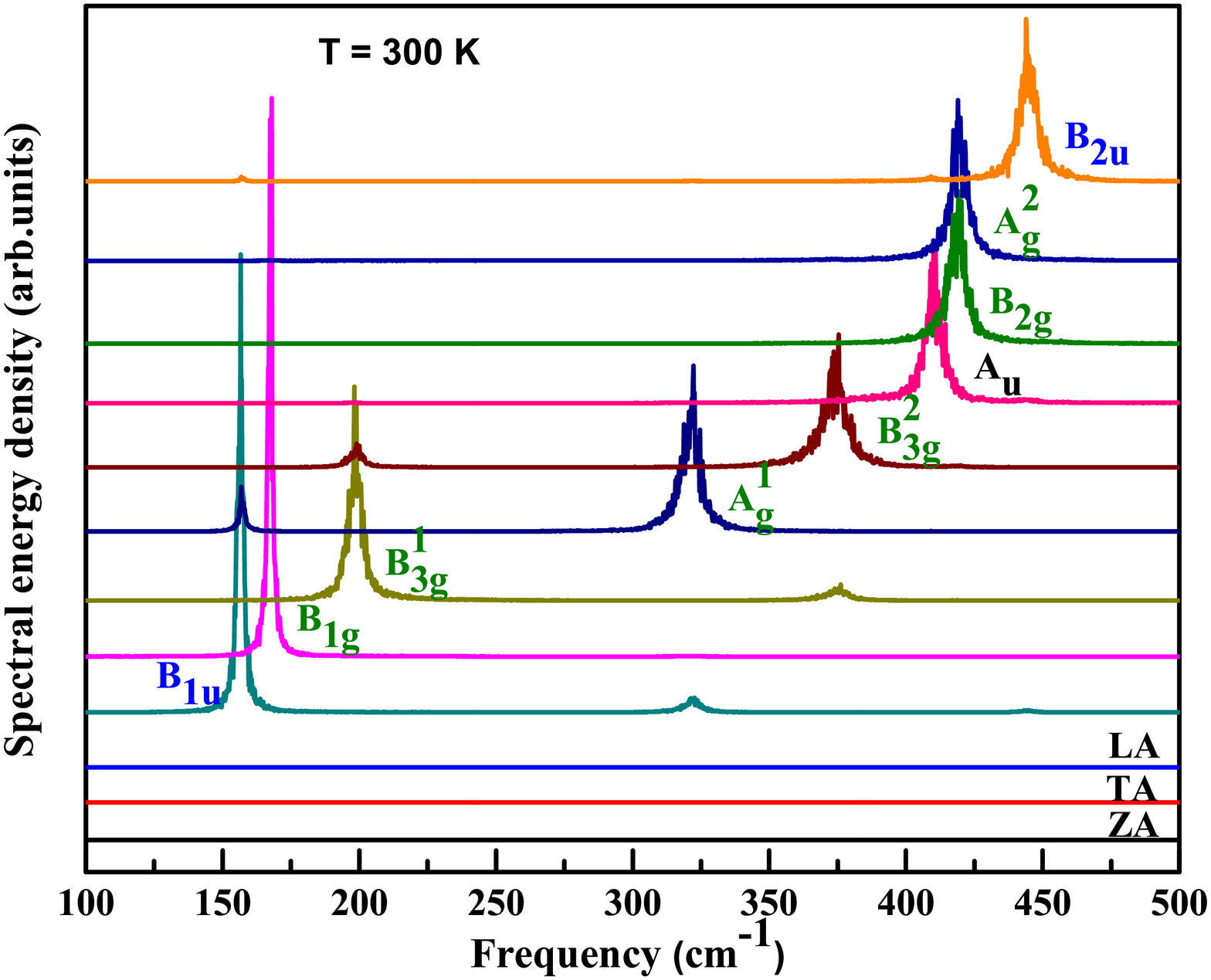}
\par\end{centering}
\centering{}\caption{(left) Phonon dispersion curve at T= 300 K; (right) mode resolved
phonon spectra at $\Gamma$ point. The IR, Raman and inactive modes
are labeled in blue, green and black color, respectively\label{fig:phonon-dispersion}.}
\end{figure*}

Finite temperature structural stability of SLBP is analyzed by computing
the phonon dispersion at 300K using SED method. Figure \ref{fig:phonon-dispersion}
shows the phonon dispersion curves at 300 K, which shares a resemblance
with LD phonon dispersion curves (Figure \ref{fig:structure-at-0K}).
The quadratic nature of ZA mode is preserved at 300 K as well. In
SED phonon dispersion all modes are real and positive, which confirms
that the SLBP sheet is dynamically stable and survives the crumpling
transitions at finite temperature. Temperature induced broadening
of phonon branches are conspicuous in SED phonon dispersion, which
is a manifestation of anharmonicity. Temperature induced broadening
is less for low lying acoustic modes than the optic modes, which indicates
that acoustic modes have higher life time and they are less prone
to temperature changes in comparison with optic modes.

\begin{figure*}
\centering{}\includegraphics[scale=0.45]{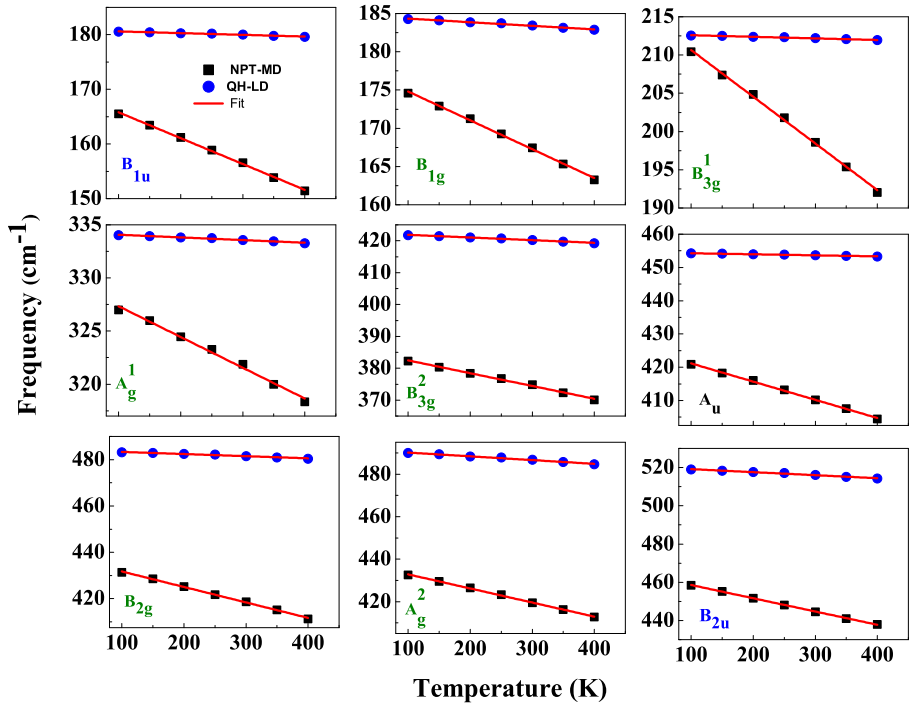}\caption{Optic mode frequency shift with temperature, blue sphere and red squares
are obtained from quasi-harmonic lattice (QH-LD) and NPT molecular
dynamics (NPT-MD) simulations\label{fig:optic-mode-shift}. The IR,
Raman and inactive modes are labeled in blue, green and black color,
respectively.}
\end{figure*}

It should be noted that many optic mode branches are merged due to
the broadening effect. Hence, in-order to analyze the vibrational
anharmonicity of optic phonons, mode resolved phonon spectra are computed
at $\Gamma$ point. Figure \ref{fig:phonon-dispersion} shows the
zone center phonon spectra at 300 K. The temperature induced peak
shift and broadening are discernible. To extract the exact mode frequency
shift, the peak positions are fitted to a Lorentzian function. Figure
\ref{fig:optic-mode-shift} shows the temperature evolution of optic
mode frequencies. In the present study, the calculations are restricted
in the temperature range of 100 K - 400 K. Above 500 K, there is an
abrupt change in potential energy, signifies that system is unstable
above this temperature. Experimentally, the decomposition temperature
of SLBP is 598 K \citep{doi:10.1063/1.4928931}, which is corroborating
with the above observation. Below 100 K the quantum effect will be
dominant, that cannot be tapped in classical MD simulations.

In-order to get a qualitative understanding on role of anharmonicity
of optic mode frequency shift, two different types of calculations
are performed. In first one, the mode frequency shift is computed
at temperature dependent lattice parameter (QH-LD). In the second
one SED method in conjunction with an isothermal-isobaric (NPT) ensemble
MD simulation (NPT-MD) is employed to obtain the frequency shift as
a function of temperature. From figure \ref{fig:optic-mode-shift},
it is evident that all QH-LD modes are red-shifted with increase in
temperature. SLBP shows thermal expansion in these temperature range
\citep{PhysRevB.92.081408}, once the lattice expands, the bonds become
softer, hence the associated force constants and then phonon frequencies
would exhibit a red-shift. The NPT-MD mode frequencies are also red-shifted
with much steeper variations. In NPT-MD, the strong phonon-phonon
coupling effects are present along with the thermal expansion. This
strong phonon-phonon coupling leads to the enhanced red-shift of mode
frequencies. This results envisages the strong anharmonicity associated
with optic phonon modes. To check the non-linearity associated with
eigen vectors, the phonon peak positions of lowest ($B_{1u}$) and
highest ($B_{2u}$) frequency modes are computed without eigen vectors.
The peak position obtained for $B_{1u}$ \& $B_{2u}$ are 158.312
cm\textsuperscript{-1} \& 446.355 cm\textsuperscript{-1}, respectively.
These values are 1.75 cm\textsuperscript{-1}and 1.73 cm\textsuperscript{-1}
higher than that obtained via projection technique, and the difference
is too small to be considered.

The temperature sensitivity of each mode is analyzed by computing
their first order temperature co-efficient $\chi$. In-order to extract
the values of $\chi$, the frequency shift of each mode is fitted
with a linear equation $\omega=\omega_{0}+\chi T$ \citep{doi:10.1063/1.4928931,C3NR02567E,doi:10.1063/1.4943546},
where $\omega_{0}$ is the extrapolated peak position to 0 K. The
$\chi$ values (Table \ref{tab:temp-sensitivity-1-1}) are deduced
from the slope of the $\omega\;vs\;T$ plot. The quasi-harmonic $\chi$
values of Raman $A_{g}^{1}$ , $B_{2g}$ and $A_{g}^{2}$ modes are
one order magnitude smaller than the NPT-MD and experimental $\chi$
values. This result again substantiate that quasi-harmonic methods
are in-adequate to capture the strong anharmonicity in SLBP. 

\begin{table}
\centering{}%
\begin{tabular}{c>{\centering}m{2cm}>{\centering}m{2cm}>{\centering}m{2cm}}
\toprule 
\multicolumn{1}{>{\centering}p{1cm}}{mode} & $\chi_{QHA}$  & $\chi_{NPT-MD}$  & expt.\tabularnewline
\midrule 
$B_{1u}$ & -0.0031 & -0.0473 & \textemdash{}\tabularnewline
\addlinespace
$B_{1g}$ & -0.0047 & -0.0377 & \textemdash{}\tabularnewline
\addlinespace
$B_{3g}^{1}$ & -0.0021 & -0.0610 & \textemdash{}\tabularnewline
\addlinespace
$A_{g}^{1}$ & -0.0025 & -0.0290 & -0.0158 \citep{doi:10.1063/1.4928931} \tabularnewline
\addlinespace
$B_{3g}^{2}$ & -0.0084 & -0.0401 & \textemdash{}\tabularnewline
\addlinespace
$A_{u}$ & -0.0031 & -0.0547 & \textemdash{}\tabularnewline
\addlinespace
$B_{2g}$ & -0.0093 & -0.0668 & -0.0315 \citep{doi:10.1063/1.4928931}\tabularnewline
\addlinespace
$A_{g}^{2}$ & -0.0176 & -0.0663 & -0.0312 \citep{doi:10.1063/1.4928931} \tabularnewline
\addlinespace
$B_{2u}$ & -0.0158 & -0.0692 & \textemdash{}\tabularnewline
\bottomrule
\addlinespace
\end{tabular}\caption{First order temperature co-efficients ($\chi)$ in cm\protect\textsuperscript{-1}/K
\label{tab:temp-sensitivity-1-1}}
\end{table}
The $\chi$ value of NPT-MD $A_{g}^{1}$ mode ( -0.0290 cm\textsuperscript{-1}/K)
shows good agreement with Raman data (-0.0158 cm\textsuperscript{-1}/K).
For other 2 Raman modes ($B_{2g}$ \& $A_{g}^{2}$) the NPT-MD $\chi$
values are roughly twice that of experimental values. This can be
understood as follows, the $A_{g}^{1}$ mode is out plane mode, while
$B_{2g}$ \& $A_{g}^{2}$ are in-plane modes. The in-plane mode frequencies
are more susceptible to substrate effect. Due to mismatch of thermal
expansion co-efficients the substrate will induce a compressive strain
on film, which alter the shift in mode frequencies, this phenomena
had already observed in graphene \citep{doi:10.1021/nl0731872,doi:10.1021/nl201488g}.
The above arguments are strengthened with the findings of Su \textit{et
al} \citep{doi:10.1063/1.4928931}, they found that the $\chi$ values
of thin BP films are more sensitive to the substrate effect than thick
BP film. Along with substrate effect, the temperature induced ripples
have significant effect on thermal expansion of 2D materials \citep{C6CP08635G},
this ripple morphology get altered while transferring to substrate,
which again affects the intrinsic vibrational properties \citep{Robinson_2015}.
The $\chi$ values of un-detected Raman active, IR and inactive modes
are also tabulated in Table \ref{tab:temp-sensitivity-1-1}. These
modes also shows significant deviation from quasi-harmonic prediction,
signifies their strong anharmonic nature. Based on the above analysis,
the temperature sensitivity of Raman active modes can be expressed
as $B_{2g}$ >$A_{g}^{2}$ >$B_{3g}^{1}$ >$B_{3g}^{2}$ > $B_{1g}$
> $A_{g}^{1}$. Fei \textit{et al} \citep{doi:10.1063/1.4894273}
reported that for $B_{2g}$ and $A_{g}^{2}$ modes the strain induced
inter-atomic distance variation is more than that of $A_{g}^{1}$,
hence they shows significant shift in mode frequencies, which is in
agreement with above observation. In the case of IR active mode, $B_{1u}$
mode shows more deviation from quasi-harmonic prediction and their
temperature sensitivity can be written as $B_{2u}$ > $B_{1u}$. 

In conclusion, the structural stability and temperature evolution
of optic phonon modes of single layer black phosphorene (SLBP) are
studied beyond the conventional quasi-harmonic methods. Finite temperature
structural stability analysis shows that SLBP sheet is thermodynamically
stable and sustain the crumpling transition. Vibrational anharmonicity
of optic modes are analyzed by performing quasi-harmonic lattice dynamics
(QH-LD) and spectral energy density (SED) method in conjunction with
an isothermal-isobaric MD (NPT-MD) simulations as a function of temperature.
Mode resolved phonon frequencies of optic modes, which includes experimentally
forbidden IR and Raman active modes are extracted. The QH-LD optic
mode frequencies shows significant deviation with respect to NPT-MD,
its ascribed to the inclusion of all higher order phonon-phonon scattering
processes in the later. Temperature sensitivity of each mode is gauged
by computing their first order temperature co-efficient, which again
exhibits the effects of strong anharmonicity associated with optic
modes in SLBP. The present study reveals that, one has to go beyond
the conventional harmonic and quasi-harmonic analysis to explain the
structural and vibrational dynamics of single layer black phosphorene,
in which anharmonicity plays crucial role at finite temperatures.

\bibliographystyle{apsrev4-1}
%

\end{document}